\documentclass{winnower}
\usepackage{subfigure}
\usepackage{amsmath}
\usepackage{amsthm}
\newcommand{\abs}[1]{\left\lvert#1\right\rvert}
\usepackage{subfigure}

\begin{document}

\title{Optimal input design for system identification using spectral decomposition}

\author{Shravan Mohan, Mithun IM and Bharath Bhikkaji\\Department of Electrical Engineering, \\Indian Institute of Technology Madras, Chennai, India}

\date{}

\maketitle

\begin{abstract}
The aim of this paper is to design a band-limited optimal input with power constraints for identifying a linear multi-input multi-output system. It is assumed that the nominal system parameters are specified. The key idea is to use the spectral decomposition theorem and write the power spectrum as $\phi_{u}(j\omega)=\frac{1}{2}H(j\omega)H^*(j\omega)$. The matrix $H(j\omega)$ is expressed in terms of a truncated basis for $\mathcal{L}^2\left(\left[-\omega_{\mbox{\tiny cut-off}},\omega_{\mbox{\tiny cut-off}}\right]\right)$. With this parameterization, the elements of the Fisher Information Matrix and the power constraints turn out to be homogeneous quadratics in the basis coefficients. The optimality criterion used are the well-known $\mathcal{D}-$optimality, $\mathcal{A}-$optimality, $\mathcal{T}-$optimality and $\mathcal{E}-$optimality. The resulting optimization problem is non-convex in general.  A lower bound on the optimum is obtained through a bi-linear formulation of the problem, while an upper bound is obtained through a convex relaxation. These bounds can be computed efficiently as the associated problems are convex.  The lower bound is used as a sub-optimal solution, the sub-optimality of which is determined by the difference in the bounds. Interestingly, the bounds match in many instances and thus, the global optimum is achieved. A discussion on the non-convexity of the optimization problem is also presented. Simulations are provided for corroboration.
\\\\
\textit{\textbf{Keywords:}} System Identification; Optimal Input Design; Spectral Decomposition; Fisher Information Matrix; Convex Optimization
\end{abstract}

\section{Introduction}
\begin{figure}[t]
\centering
\includegraphics[scale=0.25]{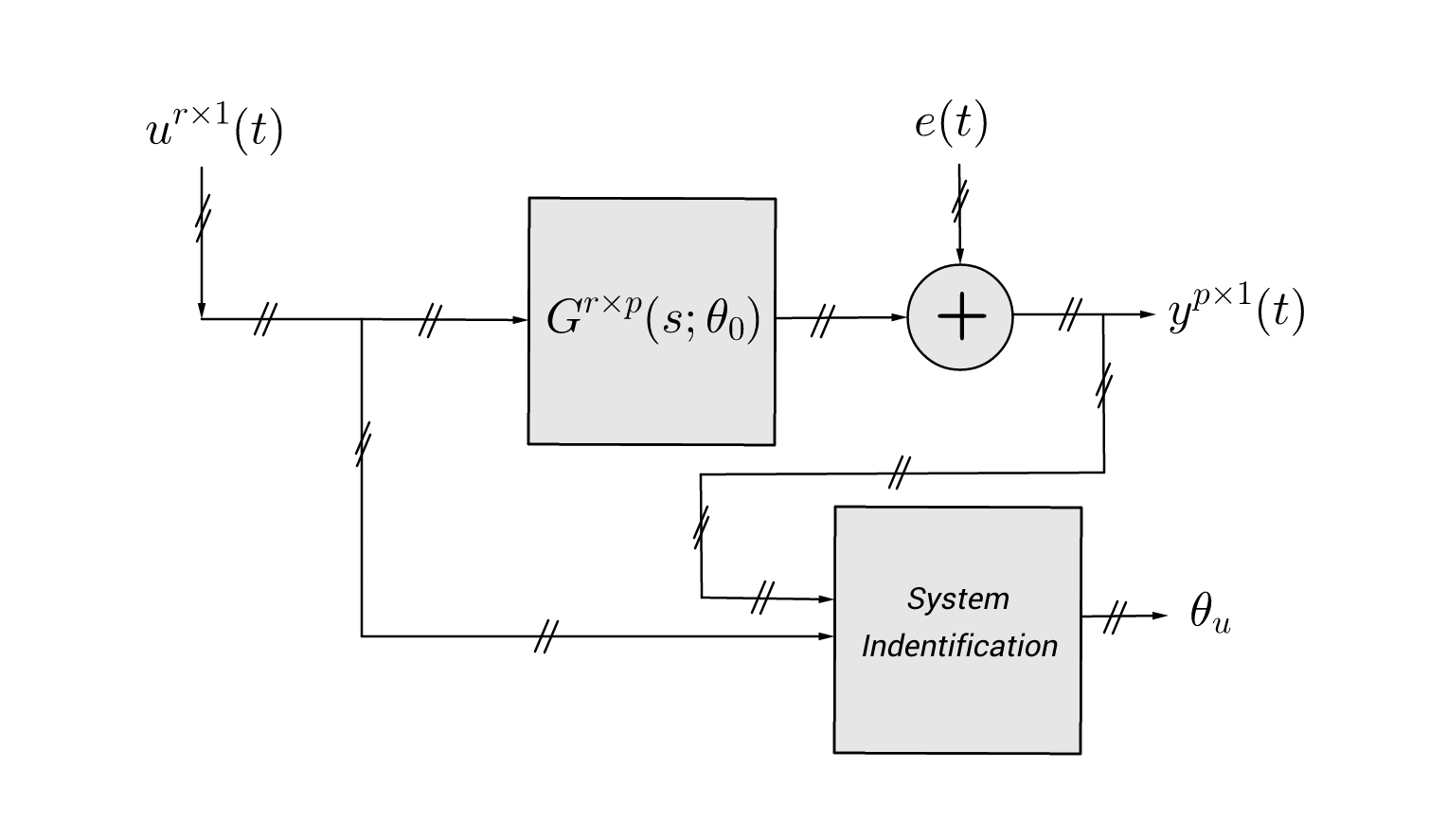}
\caption{This figure shows a general optimal input design framework. $G^{r\times p}(s;\theta_0)$ represents the linear MIMO system with nominal parameters $\theta_0$. The input vector is given by $u^{r\times 1}(t)$, $e(t)$ is a normal random vector, $i.e.$ $e(t)\sim \mathcal{N}^{p\times 1}(0,\Sigma)$, $y^{p\times 1}$ is the observed output. The estimation of the system parameters is done using subspace methods, which is represented by the box titled \textit{system identification}. The estimated system parameters is given by the set $\theta_u$. The double dashes in the figure represent multi-dimensional signals.}
\label{fig:iodiagram}
\end{figure}

In this paper, the problem of designing a band-limited optimal input for identifying a continuous time linear multi-input multi-output (MIMO) system, subject to input and output power constraints, is considered. The linear system considered is given by 
\begin{equation}
y(t)=G(s;\theta)u(t)+e(t),
\label{eq:mimo system}
\end{equation}
where
\begin{equation}
\begin{array}{l}
\displaystyle \hspace*{2cm} G_{ij}(s;\theta)=\frac{\sum_{k=0}^{m_{ij}}b_{ij}^{(k)}s^{k}}{\sum_{k=0}^{n_{ij}}a_{ij}^{(k)}s^{k}},\\
\displaystyle \hspace*{-0.5cm} \theta^{ij}=\left\{b_{ij}^{(0)},~\dots,~b_{ij}^{(m_{ij})},~ a_{ij}^{(0)},~\dots,~a_{ij}^{(n_{ij})}\right\} ~~\&~~ \theta = \bigcup_{\{i,j\}}\theta^{ij}.
\end{array}
\label{eq:mimo_transfer_function}
\end{equation}
For all $t\in \mathcal{R}^{+}$, the input $u(t)\in \mathcal{R}^{r\times 1}$, the output $y(t)\in \mathcal{R}^{p\times 1}$ and the additive white noise $e(t)\sim \mathcal{N}\left(0, \Sigma\right)$ ($\Sigma\in \mathcal{S}^+_p$, the space of positive definite $p\times p$ matrices). It is assumed that the linear system under consideration is stable and the order of elements of the matrix $G(s;\theta)$ are specified. The aim of system identification, in general, is to estimate the set of parameters $\theta$, the nominal values of which are specified, by exciting the system with inputs and recording the outputs. The schematic in Figure \ref{fig:iodiagram} shows the general system identification framework.

The quality of an unbiased estimator is judged by a suitable functional defined on the covariance matrix of the estimated system. It turns out that the covariance matrix of an unbiased estimator of a linear system is a matrix multiple of the inverse of the Fisher Information Matrix (FIM). Suppose that the set of parameters $\theta$ is indexed  and its cardinality is $N$, then the FIM for \eqref{eq:mimo system} is defined as (see \cite{kumar2015economical}):
\begin{equation}
\begin{array}{l}
\displaystyle \left[\mathcal{M}_{\theta}\right]_{l,m} =  \intop_{-\infty}^{\infty}\mbox{Tr}\left(\phi_{e}^{-1}\left(j\omega\right)\frac{\partial G(j\omega)}{\partial\theta_{l}}\phi_{u}(j\omega)\frac{\partial G(j\omega)}{\partial\theta_{m}}^*\right)d\omega,~~1\leq l,m\leq N,\\
\displaystyle \hspace*{1.8cm} \mbox{where,~~}\phi_{u}\left(j\omega\right)  =  \left[\phi_{ij}(j\omega)\right]_{\left\{1\leq i, j \leq r\right\}} \mbox{~~and~~} \phi_{e}(j\omega)=\Sigma.
\end{array}
\label{eq:M_theta}
\end{equation}
For simplicity of analysis, it is assumed that $\Sigma$ is the identity matrix. This assumption simplifies the expression of FIM to: 
\begin{equation}
\left[\mathcal{M}_{\theta}\right]_{l,m}=\intop_{-\infty}^{\infty}\mbox{Tr}\left(\frac{\partial G(j\omega)}{\partial\theta_{l}}\phi_{u}(j\omega)\frac{\partial G(j\omega)}{\partial\theta_{m}}^*\right)d\omega,~~1\leq l,m\leq N.
\label{eq:FIM_final}
\end{equation}
Since the FIM depends on the input power spectrum, it is obvious that the covariance matrix of the estimated system would depend on the input chosen for exciting the system. Therefore, the goal is to determine a power spectrum $\phi_{u}(j\omega)$ which is optimal with respect one of the following commonly used functionals defined as :
\begin{itemize}
\item $\mathcal{D}$-Optimality which maximizes the determinant of $\mathcal{M}_{\theta}$ (\cite{sanchez2012calculation,kumar2015economical}),
\item $\mathcal{A}$-Optimality which minimizes the trace of $\mathcal{M}^{-1}_{\theta}$ (\cite{aoki1970input,nahi1969optimal}),
\item $\mathcal{T}$-Optimality which maximizes the trace of $\mathcal{M}_{\theta}$ (\cite{srivastava1974comparison}), and
\item $\mathcal{E}$-Optimality which maximizes the minimum eigenvalue of $\mathcal{M}_{\theta}$
 (\cite{cheng1980optimality,constantine1981some}).
\end{itemize}
It is also natural to consider power constraints on the input signals and the output signals. The input power constraint may arise due to limitations of the excitation system. If the linear system has peak responses at some frequencies, the output power constraint would limit possible damages during excitation. The expressions for the input and the output power constraints are given by:
\begin{equation}
\begin{array}{l}
\displaystyle \intop_{-\infty}^{\infty}\mbox{Tr}\left(\phi_{u}(j\omega)\right)d\omega \leq K_{u}\mbox{~~and~~} \intop_{-\infty}^{\infty}\mbox{Tr}\left(G(j\omega)\phi_{u}(j\omega)G(j\omega)^*\right)d\omega \leq K_{y},
\end{array}
\label{eqn:io_constraints}
\end{equation}
where $K_u$ and $K_y$ are specified. It is assumed that the elements of the transfer function matrix are strictly proper. This assumption holds good in many practical systems and allows for the integral limits to be limited to a band of frequencies around zero. With the aforementioned definitions and constraints, the mathematical problem at hand is outlined in Figure \ref{fig:problem1}.
\begin{figure}[t]
\centering
\fbox{\begin{minipage}{34em}
Given $G(s;\theta)$ and $\mathcal{M}_{\theta}$ defined as in \eqref{eq:FIM_final}, find $\phi^*$ such that
\begin{equation}
\begin{array}{l}
\displaystyle \phi^* = \mbox{arg~}\max_{\phi_{u}}~\mbox{log}\left(\mbox{det}\left(\mathcal{M}_{\theta}\right)\right) \mbox{~\textit{or}~} -\mbox{Tr}\left(\mathcal{M}_{\theta}^{-1}\right) \mbox{~\textit{or}~} \sigma_{\min}\left(\mathcal{M}_{\theta}\right) \mbox{~\textit{or}~} \mbox{Tr}\left(\mathcal{M}_{\theta}\right)\\
\mbox{subject to}\\
\displaystyle \hspace*{1.5cm}\intop_{-1}^{1}\mbox{Tr}\left(\phi_{u}(j\omega)\right)d\omega \leq K_{u}\\
\displaystyle \hspace*{1.5cm}\intop_{-1}^{1}\mbox{Tr}\left(G(j\omega)\phi_{u}(j\omega)G(j\omega)^*\right)d\omega \leq K_{y}.
\end{array}
\label{eqn:optimization}
\end{equation}
Generate a signal $u(t)\in \mathcal{R}^{r\times 1}$  with a spectral density equal to $\phi_u^*$.
\end{minipage}}
\caption{The optimal input design problem.}
\label{fig:problem1}
\end{figure}

Optimal input design for linear system identification has been studied for a long time and is still an active area of research (see \cite{mehra1974optimal,  annergren2017application}). Although not comprehensive, this paragraph would touch upon some of the seminal pieces of work in this area and discuss in brief, the methodologies used therein. In each of \cite{barenthin2008identification, kumar2015economical}, the power spectrum is parameterized as 
\begin{equation}
\phi_{u}(\omega)=\sum_{k=-M}^{M}c_{k}e^{-j\omega k}.
\label{eq:power_spec_fourier_expansion}
\end{equation}
The optimal input design problem is cast as a convex optimization problem in terms of the covariance matrices $\left\{ c_{k}\right\} _{k=0}^{M}$. In particular, the Kalman-Yakubovich-Popov lemma and the Hamburger's moment condition aid the formulation. The former ensures the positive definiteness of the power spectrum while the latter guarantees the existence of a valid solution. The idea of optimal input design has many applications. \cite{ljung2011four} is  an excellent article which discusses the applications of system identification in communication systems, sensor networks and machine learning algorithms. In addition, automotive systems and chemical engineering processes also find the need for system identification and optimal input design (\cite{alberer2011identification, kumar2015economical}).

The way this paper differs from the earlier references is that the spectral density matrix is parameterized in terms of $H(j\omega)$, such that $\phi_{u}(j\omega)=\frac{1}{2}H(j\omega)H^*(j\omega)$. The matrix $H(j\omega)$ is expressed as a linear combination of a suitable truncated basis of $\mathcal{L}^2\left(-\omega_{\mbox{\small cut-off}},\omega_{\mbox{\small cut-off}}\right)$ and the coefficients become the optimization variables. Observe that this idea readily enforces positivity of $\phi_{u}(j\omega)$, which otherwise requires the Fourier basis and the Kalman-Yakubovich-Popov lemma. This also implies that the proposed method is independent of the choice of the basis. Indeed, the resulting optimization problem in this paper is non-convex in general. The optimization problem has an interesting structure which has been studied in the literature and has wide applications in signal processing and control systems. Moreover, solving this class of optimization problems still remains open.

The paper is organized as follows. The second section sets up the optimization problem, presents methods to compute lower and upper bounds, and discusses aspects related to the convexity of the problem. The third section presents simulation results, \textit{i.e.}, the input and the output waveforms, and the estimated parameters of the system. The concluding remarks are presented in the last section followed by references. The appendices to the paper present a couple of theorems and discuss the methodology of generating inputs from the optimal power spectrum. 

\section{The proposed method}
This section discusses the formulation of the optimization problem, the non-convex nature of it and a methodology for obtaining a sub-optimal solution. As mentioned earlier, the input power spectrum is limited to a pre-defined band given by $D = [-\omega_c, \omega_c]$. Now, each element of $\phi_u(j\omega)$ must belong to the space $C^{\infty}(D)$. Spectral decomposition theorem states that one can find $H(j\omega)$ belonging to $C^1(D)$ such that $\phi_u(j\omega) = H(j\omega) H^*(j\omega)$. With the knowledge that $C^1(D)$ is dense in $\mathcal{L}^2(D)$, one can as well approximate elements $H(j\omega)$ through a linear combination of a truncated basis of $\mathcal{L}^2(D)$. Therefore, each element of $H(j\omega)$ is assumed to the space of square integrable functions over $D$ denoted by $\mathcal{L}^2(D)$ (see \cite{nair2001functional}).

\subsection{Setting up the optimization problem}
As mentioned in the beginning of this section, is it assumed that each element of $H(j\omega)$ belongs to $\mathcal{L}^2(\mathcal{D})$. Thus, $H(j\omega)$ can be approximated with as a linear combination of a truncated basis for $\mathcal{L}^2(\mathcal{D})$. Suppose the truncated basis elements are given by the set $\left\{w_1, w_2, \dots, w_m\right\}$. Then $H(j\omega)$ is expressed as: 
\begin{equation}
\begin{array}{l}
\hspace{3cm} H_{m}(j\omega) =  W\mathcal{H}\\
\mbox{where,~~} W=\left[\begin{array}{cccc}
I_{r\times r} & w_1 I_{r\times r} & \dots & w_m I_{r\times r}\end{array}\right]\\
\mbox{and~~}\mathcal{H}=\left[\begin{array}{cccc}
H_{0} & H_{1} & \dots & H_{m}\end{array}\right]^{\top}, ~~ \forall i~ H_i \in \mathcal{R}^{r\times r}.\label{eq:mathcalH}
\end{array}
\end{equation}
The matrix $\mathcal{H}$ comprises of the coefficients which need to be determined by solving an optimization problem. As a consequence, the input power spectral density can be expressed as:
\begin{equation}
\phi_{u}^{m}(j\omega)=\frac{1}{2}W(j\omega)\mathcal{H}\mathcal{H}^{\top}W(j\omega)^*.\label{eq:approx_power_spec_density}
\end{equation}
Using \eqref{eq:approx_power_spec_density}, the constraint on the total input power is given by:
\begin{equation}
\intop_{-\infty}^{\infty}\mbox{Tr}\left(\frac{1}{2}W(j\omega)\mathcal{H}\mathcal{H}^{\top}W(j\omega)^*\right)d\omega\le K_{u}.\label{eq:inp_cons_para_HHT}
\end{equation}
Interchanging the integral operator and the trace operator, one obtains 
\begin{equation}
\begin{array}{l}
\intop_{-\infty}^{\infty}\mbox{Tr}\left(\frac{1}{2}W(j\omega)\mathcal{H}\mathcal{H}^{\top}W(j\omega)^*\right)d\omega  =  \mbox{Tr}\left(\mathcal{H}^{\top}\left(\intop_{-1}^{1}\frac{1}{2}W(j\omega)^*W(j\omega)d\omega\right)\mathcal{H}\right) \\
\hspace{5.5cm} =  \mbox{Tr}\left(\mathcal{H}^{\top}\mathcal{W}\mathcal{H}\right)\\
\hspace{5.5cm} = h^{\top}P_{I}h,\\
\hspace*{2cm} \mbox{where,~~} h=\mbox{vec}(\mathcal{H}) \mbox{~~and~~} 
P_{I}=\left[\begin{array}{cccc}
\mathcal{W}\\
 &  \ddots\\
 &  &  \mathcal{W}
\end{array}\right].
\label{eq:lhs_inp_pow_form}
\end{array}
\end{equation}
Since $\mathcal{W}$ is a positive definite matrix by (\ref{eq:mathcalW_PD}), $P_{I}$ is also  positive definite. In short, the total input power constraint given by \eqref{eq:inp_cons_para_HHT} is equivalent to
\begin{equation}
h^{\top}P_{I}h \leq K_u.
\end{equation}
Note that $h\in \mathcal{R}^{mr^2}$. Similarly the total output power constraint given by
\begin{equation}
\intop_{-\infty}^{\infty}\mbox{Tr}\left(G(j\omega)W(j\omega)\mathcal{H}\mathcal{H}^{\top}W(j\omega)^*G(j\omega)^*\right)d\omega\le K_{y}-p.\label{eq:out_cons_para_HHT}
\end{equation}
can be rewritten as
\begin{equation}
\begin{array}{l} 
\hspace*{5cm} \displaystyle h^{\top}P_{O}h \leq K_y, \mbox{~where} \\
\displaystyle P_{O}=\left[\begin{array}{cccc}
\mathcal{W}_{G}\\
 &  \ddots\\
 &  &  \mathcal{W}_{G}
\end{array}\right] 
\displaystyle \mbox{~~and~~} \mathcal{W}_{G}=\frac{1}{2}\intop_{-1}^{1}W(j\omega)^*G(j\omega)^*G(j\omega)W(j\omega)d\omega.\label{eq:mathcalW_G}
\end{array}
\end{equation}
Note that, as in (\ref{eq:mathcalW_PD}), it can be shown that $\mathcal{W}_{G}$
is positive definite and so is $P_O$. Finally, the FIM $\mathcal{M}_{\theta}$ can be rewritten as
\begin{equation}
\begin{array}{l}
\displaystyle \left[\mathcal{M}_{\theta}\right]_{l,m}=\intop_{-\infty}^{\infty}\mbox{Tr}\left(\frac{\partial G(j\omega)}{\partial\theta_{l}}\phi_{u}(j\omega)\frac{\partial G(j\omega)}{\partial\theta_{m}}^*\right)d\omega = \mbox{Tr}\left(\left(\mathcal{H}^{\top}\tilde{G}_l\right)\left(\mathcal{H}^{\top}\tilde{G}_m\right)^{\top}\right),
\label{eq:FIM_final}
\end{array}
\end{equation}
where $\displaystyle \tilde{G}_l = \intop_{-\infty}^{\infty}\frac{\partial G(j\omega)}{\partial\theta_{l}}W$. In other words, every element of the FIM is a quadratic in the coefficients $h$. Now assuming $\displaystyle \Phi(.) = \mbox{logdet}(.)$ or $\mbox{$\sigma_{\min}$}(.)$ or $-\mbox{trace-inv}(.)$ or $\mbox{trace}(.)$, the optimization problem \eqref{eqn:optimization} is equivalent to 
\begin{equation}
\begin{array}{l}
\displaystyle \max_{h\in \mathcal{R}^{mr}} ~\Phi(\mathcal{M}_{\theta}(h))\\
\displaystyle \mbox{subject to}\\
\displaystyle \hspace*{1cm} h^{\top} P_Ih \leq K_u\\
\displaystyle \hspace*{1cm} h^{\top} P_Oh \leq K_y.
\end{array}
\label{eq:optim1}
\end{equation}

\subsection{Finding a suboptimal solution}
To the best of the authors' knowledge, the optimization problem \eqref{eq:optim1} does not have an efficient algorithm. The problem, due to its quadratic structure, renders itself to bi-linear forms, as well as semi-definite convex relaxations. The method proposed here depends on these derived forms and has two steps in it: (i) find a lower bound to \eqref{eq:optim1} and (ii) find an upper bound to \eqref{eq:optim1}. The first step, which obtains a lower bound to the optimization problem through a bi-linear form, is iterative and in each iteration, a convex problem is solved. The procedure is asymptotically convergent and produces a feasible solution. The second step find an upper bound to the optimal value of \eqref{eq:optim1} through a convex relaxation. Hence, it can be computed efficiently with semi-definite programming. The solution corresponding to the lower bound is a used as a sub-optimal solution to \eqref{eq:optim1}. The sub-optimality of the solution is determined by the difference between the upper and lower bounds. 
\\ \\
\textit{\underline{Lower bound}}: To obtain a lower bound, the quadratic form in each of the elements of $\mathcal{M}_{\theta}$ is converted into a bi-linear form. The input and output power constraints, which are convex, are retained as they are. In other words, 
\begin{equation}
\begin{array}{l}
x^{\top}M_{i,j}x \mbox{~~is converted to~~} x^{\top}M_{i,j}y \mbox{~~for all $i,j$}.
\end{array}
\end{equation}
The bi-linear optimization problem is described, for example, in \cite{konno1976maximization}. Note that once one of the variables ($x$ or $y$) is fixed to some vector, \eqref{eq:optim1} becomes a convex optimization problem in the other variable. Starting from an arbitrary feasible point $y=y_0$, the convex optimization problem in $x$ is solved to obtain $x^*$. For the next iteration, $y$ is fixed to the vector the $x^*$. The resulting vector of each iteration acts as a lower bound  to \eqref{eq:optim1} since it is a feasible solution. Note that with each iteration, the value of the cost function can only increase and hence, the lower bound becomes tighter. The process asymptotically converge to a value but however, a discussion on the rate of convergence is out of the scope of this paper. The algorithm, also termed as the `hill-climbing' procedure, is outlined in  Figure~\ref{fig:lower_bound}.
\begin{figure}[t]
\centering
\fbox{\begin{minipage}{32em}
$y=y_0$\\ \\
while(count $\leq$ 100 $||$ tol $\leq$ 1e-6) do
\begin{equation}
\begin{array}{l}
\displaystyle \max_{x\in \mathcal{R}^N} ~\Phi(\mathcal{M}(x,y)),~~\mbox{where~} \mathcal{M}_{ij}(x,y) = x^{\top} M_{i,j}y \mbox{~and~} M_{i,j}\in \mathcal{R}^{N\times N}\\
\hspace*{3cm}~\&~ \Phi(.) = \mbox{logdet}(.)|\mbox{$\sigma_{\min}$}(.)|\mbox{-trace-inv}(.)|\mbox{trace}(.)\\
\displaystyle \mbox{subject to}\\
\displaystyle \hspace*{2cm} x^{\top} P_Ix \leq 1\\
\displaystyle \hspace*{2cm} x^{\top} P_Ox \leq 1 \\ \\
\displaystyle \mbox{tol} = ||x-y||_2\\
\displaystyle y = x
\end{array}
\end{equation}
end
\end{minipage}}
\caption{Lower bound algorithm: $y_0$ is a random initial vector, the maximum iteration count is set to 100 and the tolerance bound for convergence is set to 1e-6. $\mathcal{M}(x,y)$ represents the bi-linear form of the Fisher Information Matrix, $P_I$ represents the total input power matrix and $P_O$, the total output power matrix. The functional $\Phi$ can be chosen either of the following functions: $\mbox{logdet}(.)$ or $\mbox{$\sigma_{\min}$}(.)$ or $\mbox{-trace-inv}(.)$ or $\mbox{trace}(.)$.}
\label{fig:lower_bound}
\end{figure}
\\ \\
\textit{\underline{Upper bound}}: The upper bound is determined by a standard semi-definite convex relaxation of \eqref{eq:optim1}. It can be shown that any homogeneous quadratic $x'Ax$ can also be written as (see \cite{boyd2004convex})
\begin{equation}
\begin{array}{l}
x'Ax = \mbox{Tr}(AX), \mbox{~~where~~} X = xx^{\top}.
\end{array}
\end{equation}
This transformation implies that \eqref{eq:optim1} is equivalent to 
\begin{equation}
\begin{array}{l}
\displaystyle \max_{X\in \mathcal{R}^{N\times N}} ~\Phi(\mathcal{M}_{\theta}(X)),~~\mbox{where~} \left[\mathcal{M}_{\theta}\right]_{ij}(X) = \mbox{Tr}(A^{ij}X) \\
\displaystyle \mbox{subject to}\\
\displaystyle \hspace*{1cm} \mbox{trace}(P_IX) \leq 1\\
\displaystyle \hspace*{1cm} \mbox{trace}(P_OX) \leq 1 \\
\displaystyle \hspace*{1cm} X \succeq 0 \mbox{~~and~~} \mbox{Rank}(X) = 1
\end{array}
\label{eq:optim2}
\end{equation}
Dropping the Rank-1 in \eqref{eq:optim2} results in a convex semi-definite optimization problem shown in Figure~\ref{fig:upper_bound}. Being a relaxation of the original optimization problem, the solution to \eqref{eq:optim2} gives an upper bound to \eqref{eq:optim1}. Moreover, the bound can be obtained efficiently as the associated problem is convex. However, unlike the lower bound algorithm, the solution matrix $X^*$ need not be have a rank of 1, in which case deriving a feasible solution for \eqref{eq:optim1} from it may not be possible.
\\ \\
\underline{\textit{Remark}}: It might be necessary to have more power constraints in the optimization problem. In particular, it might be necessary to constrain input and output powers at the individual ports, in addition to the total power constraints. Moreover, the power constraints may also be non-convex, particularly when the constraint is to deliver a minimum amount of input or output power. In all the cases mentioned here, the algorithm for the upper bound does not change, except for the increase in the number of constraints. The lower bound algorithm would need a few changes. Firstly, the constraints are also changed to their corresponding bi-linear forms. Secondly, the step ``$y=x$" in Figure \ref{fig:lower_bound} needs to be replaced by ``$y=\frac{x_k+x_{k-1}}{2}$". In other words, $y$ is replaced by the average of the resulting vectors of the last two iterates.

\begin{figure}[t]
\centering
\fbox{\begin{minipage}{32em}
\begin{equation}
\begin{array}{l}
\displaystyle \max_{X\in \mathcal{R}^{N\times N}} ~\Phi(\mathcal{M}(X)),~~\mbox{where~} \mathcal{M}_{ij}(X) = \mbox{trace}(M_{i,j}X),~~ M_{i,j}\in \mathcal{R}^{N\times N}\\
\hspace*{3cm}~\&~ \Phi(.) = \mbox{logdet}(.)|\mbox{$\sigma_{\min}$}(.)|\mbox{trace-inv}(.)|\mbox{trace}(.)\\
\displaystyle \mbox{subject to}\\
\displaystyle \hspace*{2cm} \mbox{Tr}(P_IX) \leq 1\\
\displaystyle \hspace*{2cm} \mbox{Tr}(P_OX) \leq 1 \\
\displaystyle \hspace*{2cm} X \succeq 0
\end{array}
\end{equation}
\end{minipage}}
\caption{Upper bound algorithm: $\mathcal{M}(X)$ represents the relaxed form of the Fisher Information Matrix, $P_I$ represents the total input power matrix and $P_O$, the total output power matrix. The functional $\Phi$ can be chosen either of the following functions: $\mbox{logdet}(.)$ or $\mbox{$\sigma_{\min}$}(.)$ or $\mbox{-trace-inv}(.)$ or $\mbox{trace}(.)$.}
\label{fig:upper_bound}
\end{figure}

\subsection{A discussion on quadratic maps}
This subsection discusses the convexity properties of the optimization problem given in \eqref{eq:optim1}. As mentioned earlier, every element of the FIM $\mathcal{M}_{\theta}(h)$ is a homogeneous quadratic in $h$. For clarity, the optimization problem is written in an expanded form as:
\begin{equation}
\begin{array}{l}
\displaystyle \max_{h\in \mathcal{R}^{mr^2}} ~~\Phi\left(
\left[\begin{array}{cccc}
h^{\top}M_{1,1}h &  \dots  &h^{\top}M_{1,N}h \\
\vdots & & \vdots\\
h^{\top}M_{N,1}h &  \dots &h^{\top}M_{N,N}h \\
\end{array}\right]\right) \\
\displaystyle \mbox{subject to}\\
\displaystyle \hspace*{2cm} h^{\top} P_Ih \leq K_u\\
\displaystyle \hspace*{2cm} h^{\top} P_Oh \leq K_y, \\
\end{array}
\end{equation}
where $M_{i,j}$'s are symmetric matrix of dimension $mr\times mr$. Moreover, by construction, the symmetric matrices are such that the FIM in the argument of $\Phi$ is positive semidefinite for all $h\in \mathcal{R}^{mr}$. The convexity of this optimization problem depends on the convexity of the image of $\mathcal{R}^{mr}$ under the quadratic map $\displaystyle Q: \mathcal{R}^{mr^2} \rightarrow \mathcal{R}^{\frac{N(N+1)}{2}+2}$ given by:
\begin{equation}
\begin{array}{l}
Q(x) = \left[
x^{\top}M_{1,1}x ~\dots~ x^{\top}M_{i,j}x ~\dots~ x^{\top}M_{N,N}x ~~x^{\top}P_Ix ~~x^{\top}P_Ox
\right]^{\top};~~ i\leq j.
\end{array}
\label{eq:quad_map1}
\end{equation}
It is assumed here that the total number of coefficients given by $mr$ is much larger than the number of parameters to be estimated given by $N$. Quadratic maps and the convexity of its image have been studied earlier in literature and some of the important results are pointed out here.

\cite{ramana1994quadratic} were one of the first few to study, in detail, the convexity of images under quadratic transformations. They give necessary and sufficient condition for characterizing such maps. However, the condition are not checkable for general cases. \cite{polyak1998convexity} proved that the image of the unit disk under any ``definite" quadratic map $Q:\mathcal{R}^n\rightarrow \mathcal{R}^3$ is convex for $n\geq 3$. \cite{sheriff2013convexity} presented results for ``stable convexity" and discussed its applications in analyzing coupled systems. \cite{dymarsky2014convexity} showed the convexity of the image of an open ball in $\mathcal{R}^n$ under quadratic maps. Since the dimension of the range space in this paper is much larger than 3, these results cannot be applied here. \cite{beck2009convexity} presented some startling generalizations of earlier results on the subject. One of the results can be applied to obtain a convex relaxation of the optimization  problem in this paper, but only at the cost of a large number of redundant variables.  \cite{hiriart2002permanently} provided an excellent overview on the convexity aspects of the images of quadratic maps in their paper. These references point out that quadratic maps with convex images are rather exceptions than the rule. 

\subsection{Input generation and system identification}
The methodology for generating inputs real inputs with the optimal power spectrum obtained from the optimization procedure is outlined in Appendix B. The methodology essentially discretizes the power spectrum, concatenates it with a random phase and performs inverse Fourier transform to obtain the inputs. The system is excited with the the designed optimal inputs and the outputs are recorded. The system identification is done using the Instrument Variables method, which is an unbiased estimator of the system parameters. For details on the system identification algorithm, the reader is directed to  \cite{soderstrom2002instrumental}. 

\section{Simulations}
\begin{figure}[t]
\centering
\includegraphics[width=6.5in]{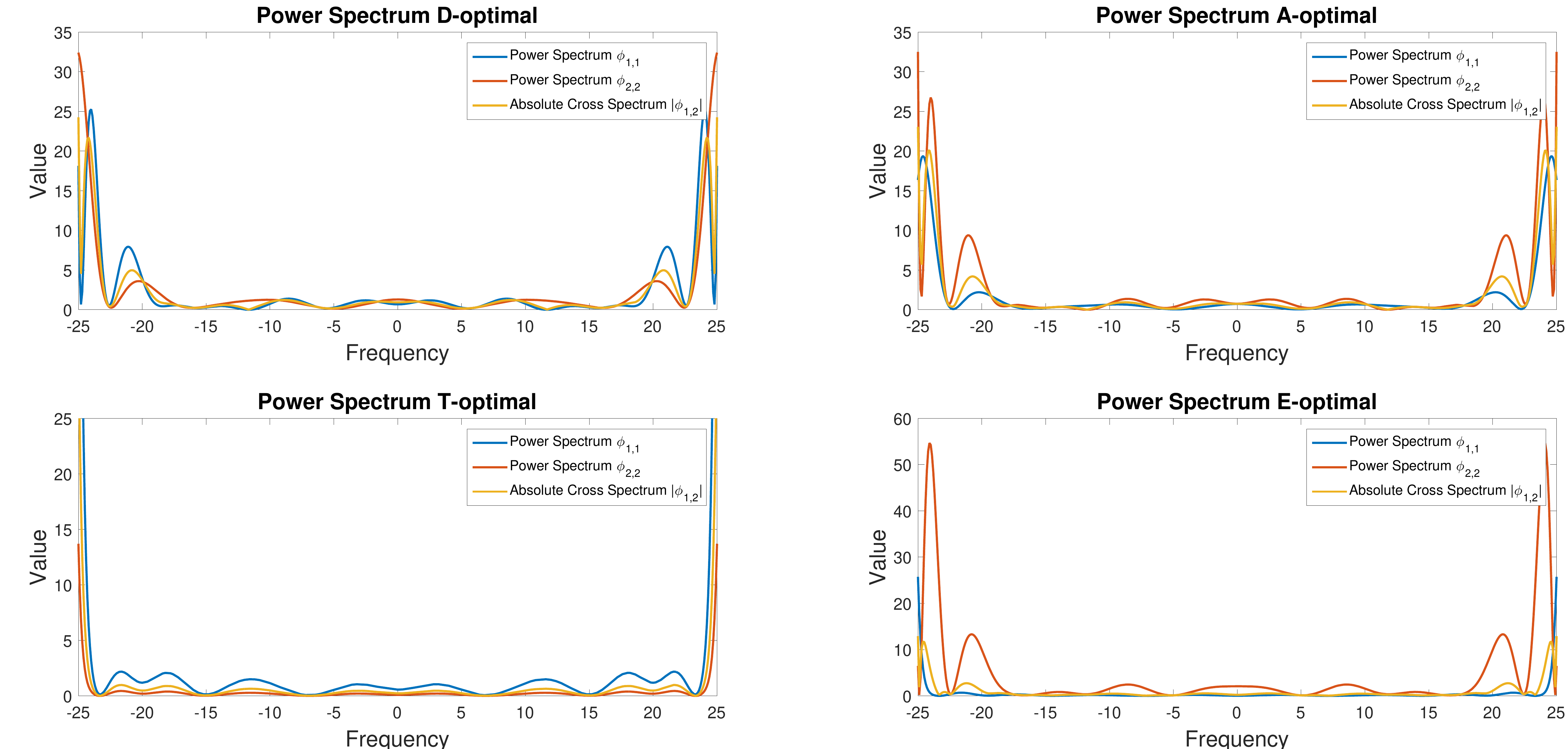}
\caption{This figure shows the optimal power spectrum obtained using the proposed method for the four different optimality criterion. The calculations were done by parameterizing $H(j\omega)$ with Chevyshev's polynomials upto the $13^{\rm{th}}$ order. The absolute value of the cross-spectrum is also plotted.}
\label{fig:spec}
\end{figure}

\begin{figure}[h!]
\centering
\includegraphics[width=6.5in]{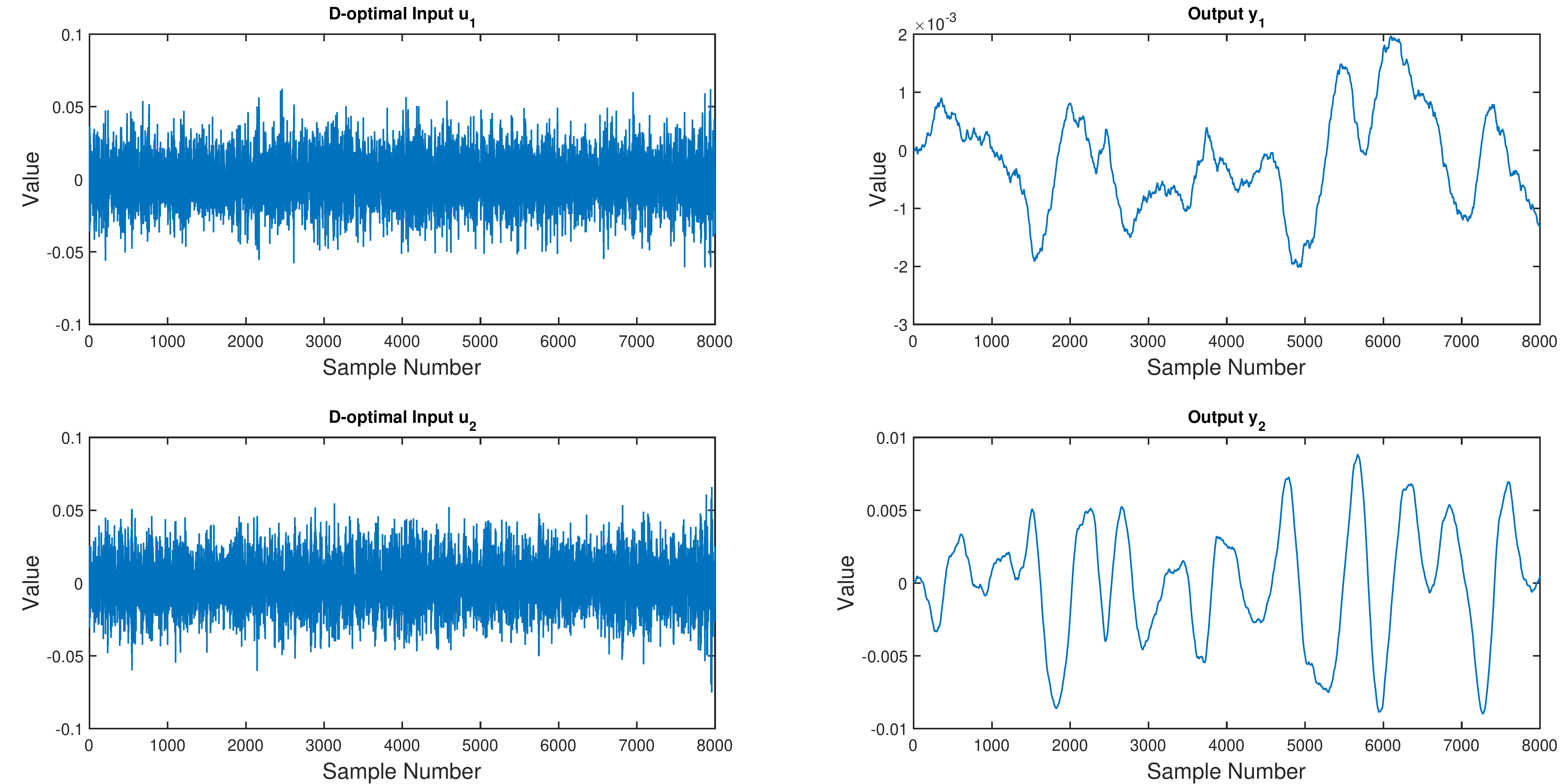}
\caption{The inputs corresponding to the $\mathcal{D}$-optimal spectrum are shown on the left. The outputs obtained by exciting the system with the inputs are shown to the right.}
\label{fig:iod}
\end{figure}

\begin{figure}[h!]
\centering
\includegraphics[width=6.5in]{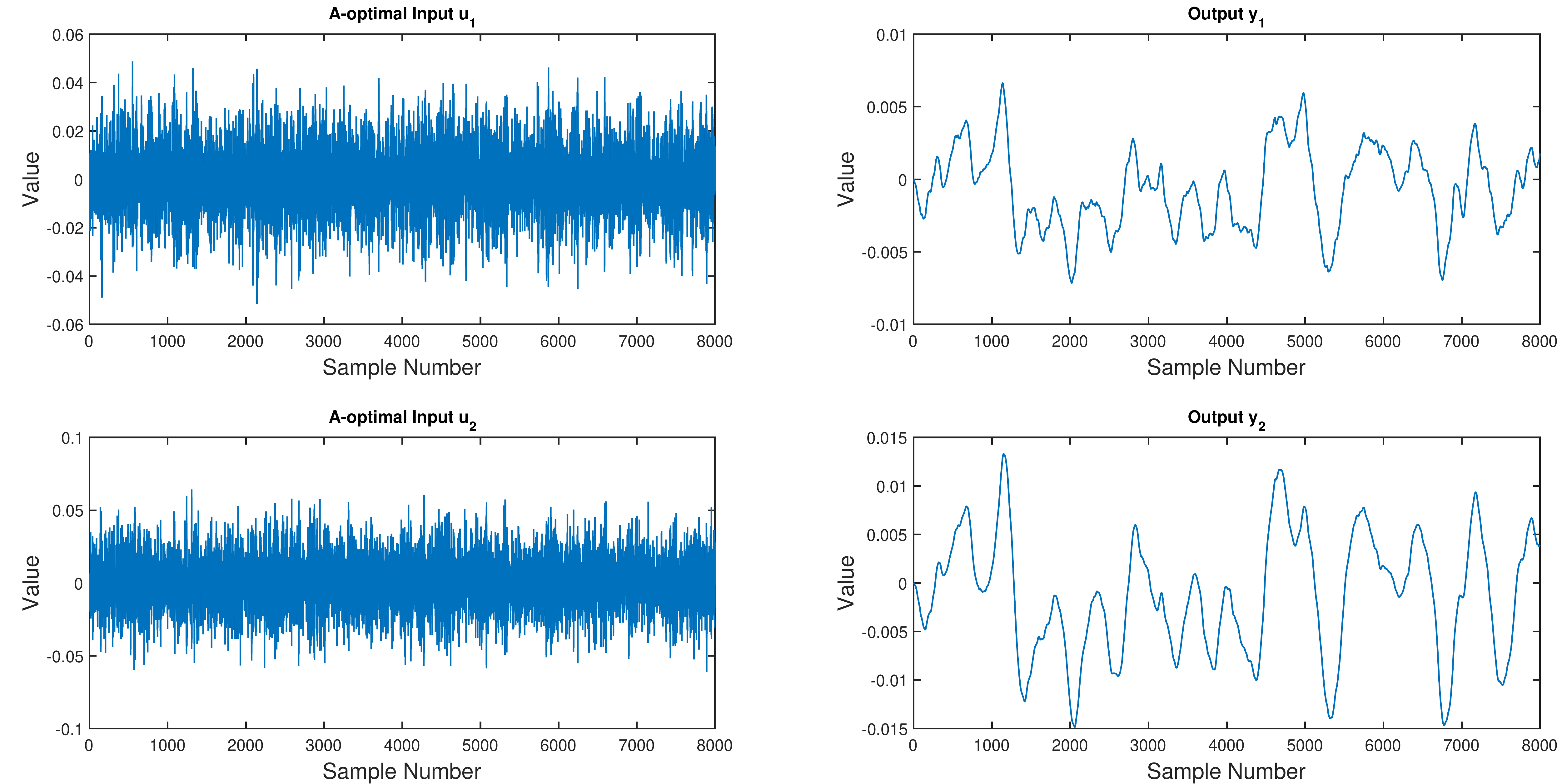}
\caption{The inputs corresponding to the $\mathcal{A}$-optimal spectrum are shown on the left. The outputs obtained by exciting the system with the inputs are shown to the right.}
\label{fig:ioa}
\end{figure}

\begin{figure}[h!]
\centering
\includegraphics[width=6.5in]{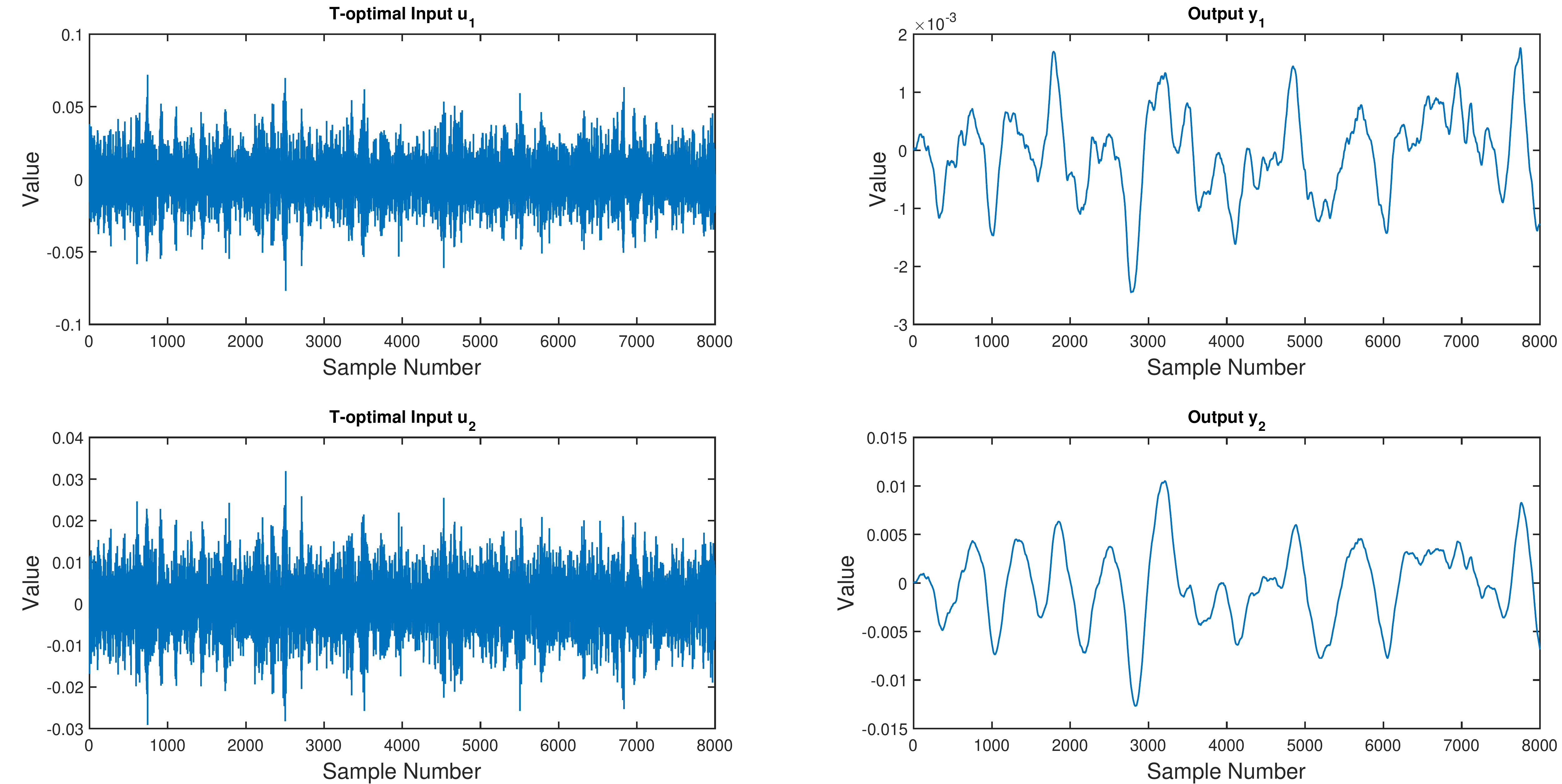}
\caption{The inputs corresponding to the $\mathcal{T}$-optimal spectrum are shown on the left. The outputs obtained by exciting the system with the inputs are shown to the right.}
\label{fig:iot}
\end{figure}

In most practical applications, a system is fabricated with a desired/nominal set of parameters. But fabrication is seldom perfect and therefore, the actual parameters vary to a small extent around the desired locations. For optimal input design, the nominal system parameters are used for generating the optimal inputs, while the system with actual parameters generates the outputs when actuated with the optimal inputs. For simulations, a 2-input and 2-output system was chosen. The nominal transfer function matrix and the actual transfer function matrix are given by:
\[
\displaystyle 
  \underbrace{\begin{bmatrix} \displaystyle
    \displaystyle\frac{3.2}{s^2 + 1.6 s + 1.5}  &\displaystyle\frac{3.1  }{  s^2 + 1.9 s + 1} \\ \\  
\displaystyle\frac{5.2 }{s^2 + 0.5 s + 0.95}  & \displaystyle\frac{1.5  }{ s^2 + 0.9 s + 0.3}
  \end{bmatrix}}_{G},~
  \displaystyle 
  \underbrace{\begin{bmatrix} \displaystyle
    \frac{3.125}{s^2 + 1.5 s + 1.563} & \displaystyle \frac{3.188}{s^2 + 2 s + 1.063} \\ \\
    \displaystyle \frac{5.313}{s^2 + 0.5 s + 1.063} & \displaystyle \frac{ 1.563}{s^2 + s + 0.312}
  \end{bmatrix}}_{G_{nom}}.
\]
The total input power and the total output power were constrained to a maximum of 1000 units. Chebyshev's polynomials up to the $13^{\rm{th}}$ order  were considered for parameterizing $H(j\omega)$. It was observed that lower bounds and the upper bounds matched in the $\mathcal{D}$, $\mathcal{A}$ and $\mathcal{E}$ optimality criterion, while they were off by less than $0.1\%$ in case of $\mathcal{T}$ optimality. It has been observed that the lower and upper bounds match in several instances. However, there is no mathematical proof for the observation yet. The optimal power spectrums obtained using the proposed method have been depicted in Figure \ref{fig:spec}. The inputs generated according to the optimal spectrums corresponding to the four different optimality criterion are shown in Figures \ref{fig:iod}, \ref{fig:ioa}, \ref{fig:iot}. In addition, the corresponding outputs, corrupted with uncorrelated additive Gaussian white noise vectors, are also  shown in Figures \ref{fig:iod}, \ref{fig:ioa}, \ref{fig:iot}. The standard deviation of the measurement noise is assumed to be equal to one-tenth of the  minimum of the root-mean-squared values of the outputs. The system identification was done using Instrument Variables method and the results corresponding to the different optimality criteria are given below.
\[
\displaystyle 
  \underbrace{\begin{bmatrix} \displaystyle
\displaystyle\frac{3.21 }{s^2 + 1.603 s + 1.497} &\displaystyle\frac{3.105 }{s^2 + 1.897 s + 0.998} \\ \\  
\displaystyle\frac{ 5.073 }{s^2 + 0.492 s + 0.951} & \displaystyle\frac{ 1.387 }{s^2 + 0.856 s + 0.284}
  \end{bmatrix}}_{G_{\mathcal{D}}},~
  \displaystyle 
  \underbrace{\begin{bmatrix} \displaystyle
\displaystyle\frac{3.555 }{s^2 + 1.639 s + 1.501} & \displaystyle\frac{2.742 }{s^2 + 1.863 s + 0.961} \\ \\ 
\displaystyle\frac{ 5.252 }{s^2 + 0.502 s + 0.952} & \displaystyle\frac{1.458 }{s^2 + 0.884 s + 0.295}
  \end{bmatrix}}_{G_{\mathcal{A}}},
\]

\[
\displaystyle 
  \underbrace{\begin{bmatrix} \displaystyle
\displaystyle\frac{3.197}{s^2 + 1.606 s + 1.506} & \displaystyle\frac{3.082}{s^2 + 1.920 s + 0.999} \\ \\ 
\displaystyle\frac{ 5.172}{s^2 + 0.498 s + 0.951} & \displaystyle\frac{ 1.425 }{s^2 + 0.873 s + 0.292}
  \end{bmatrix}}_{G_{\mathcal{T}}},~
  \displaystyle 
  \underbrace{\begin{bmatrix} \displaystyle
\displaystyle\frac{2.787 }{s^2 + 1.502 s + 1.488} & \displaystyle\frac{3.144 }{s^2 + 1.905 s + 1.006} \\ \\ 
\displaystyle\frac{ 5.308 }{s^2 + 0.499 s + 0.952} & \displaystyle\frac{ 1.473 }{s^2 + 0.887 s + 0.296}
  \end{bmatrix}}_{G_{\mathcal{E}}}.
\]

\begin{figure}[h!]
\centering
\includegraphics[width=6.5in]{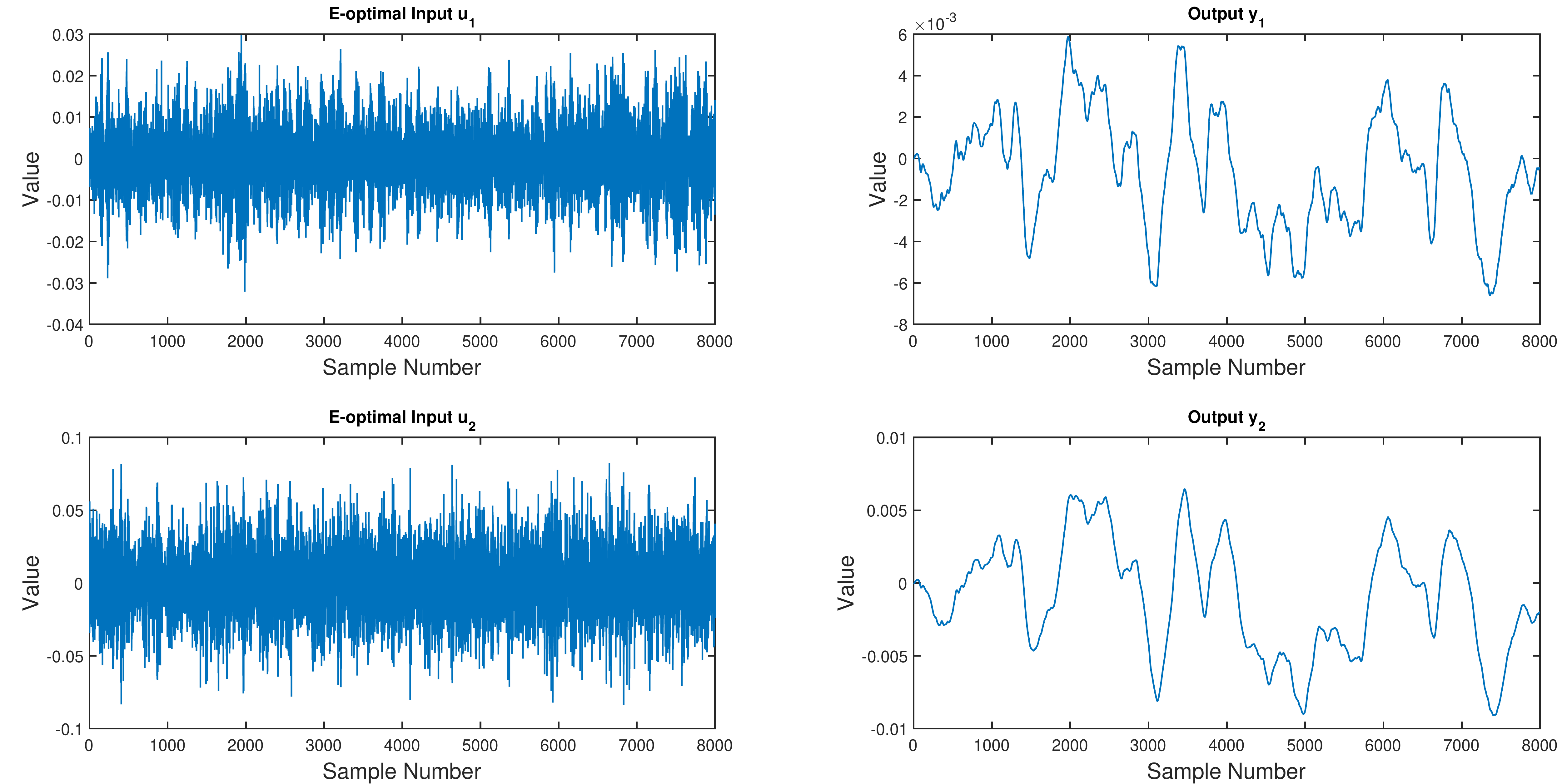}
\caption{The inputs corresponding to the $\mathcal{E}$-optimal spectrum are shown on the left. The outputs obtained by exciting the system with the inputs are shown to the right.}
\label{fig:ioe}
\end{figure}

\section{Conclusion}
In this paper, a distinct paradigm to design a band-limited optimal inputs for identifying a linear continuous time MIMO system was presented. Unlike conventional schemes where the power spectral density is parametrized with covariance matrices, this paper used spectral decomposition. The input spectra $\phi_{u}(j\omega)$ was factored as $\frac{1}{2}H(j\omega)H^*(j\omega)$ and $H(j\omega)$ was approximated by a truncated basis for $\mathcal{L}^2([-\omega_c, \omega_c])$, where $\omega_c$ is the cut-off frequency. The coefficients of the finite linear combination thus became the optimization parameters. The parameterization ensured the positivity of the power spectral density matrix, irrespective of the basis chosen for $\mathcal{L}^2(\mathcal{D})$. The optimization problem of determining the optimal coefficients for $\mathcal{D},~ \mathcal{A}$, $\mathcal{T}$ or $\mathcal{E}$ optimality criteria turned out to be a non-standard optimization problem. A discussion on the convexity aspect of the problem was also presented. A lower bound for the optimal value was obtained by first converting the optimization problem into a bi-linear form and then solving it using an iterative procedure. In each of the iterations, a convex problem was solved which made the algorithm computationally efficient. An upper bound for the optimal value was obtained by first converting the optimization problem into a rank constrained semi-definite problem. The Rank-1 constraint was then dropped and the resulting convex semi-definite program was solved using standard routines. If the upper bound turned out to be a Rank-1 solution, the vector defining its column space was chosen as the optimal solution; else the solution corresponding to the lower bound was chosen. The sub-optimality of the final solution was determined by the difference in the bounds. Interestingly, the bounds matched in many instances of simulations. With the optimal spectral density, the corresponding inputs were generated. The linear MIMO system was stimulated with these inputs and the outputs, corrupted with measurement noise, were recorded. The  Instrument Variables method was used to estimate the system parameters. The estimated system was observed to be in close agreement with the original system.

As for future directions of work, several interesting questions can be posed. The authors would like to point to a few: (i) The choice of basis for $\mathcal{L}^2(\mathcal{D})$ and its effect on the obtained optimality, and (ii) the ratio of the lower bound to the upper bound, in general.

\bibliographystyle{abbrvnat}

\appendix
\section{Theorems and proofs}
\textit{\textbf{Theorem.}} Suppose $\mathcal{W} = \frac{1}{2}\intop_{-1}^{1}W(j\omega)^*W(j\omega)d\omega$. Then $\mathcal{W}$ is an $r(m+1)\times r(m+1)$ positive definite matrix.
\begin{proof}
It is clear that $\mathcal{W}$ is positive semidefinite. Suppose for some $x\ne0$
\begin{eqnarray}
0 ~=~ x^*\mathcal{W}x & = & \intop_{0}^{1}x^*W(j\omega)^*W(j\omega)xd\omega,\nonumber \\
 & \Rightarrow & x^*W(j\omega)^*W(j\omega)x=0,\forall\omega\nonumber \\
 & \Rightarrow & W(j\omega)x=0,\,\forall\omega\nonumber \\
 & \Rightarrow & x=0, ~ \mbox{a contradiction}. \nonumber \label{eq:mathcalW_PD}
\end{eqnarray}
\end{proof}
\hspace*{-0.55cm} \textit{\textbf{Theorem.}} Suppose $\Phi$ is a matrix function defined on $[-\pi, \pi]$ such that (i) $\Phi(\theta)\in \mathbb{C}^{q\times q}$ is a Hermitian nonnegative definite matrix for all $\theta \in [-\pi, \pi]$, (ii) $\Phi(\theta) = \Phi^{\top}(-\theta)$ and (iii) $\Phi$ is integrable and admits a Fourier series. Then there exists a matrix function $\Psi$ defined on $[-\pi, \pi]$ such that $\Phi(\theta) = \Psi(\theta)\Psi^*(\theta)$. Moreover, $\Psi$ admits a Fourier series. 
\begin{proof}
See \cite{wilson1972factorization}.
\end{proof}
\section{Design of inputs from optimal power spectrum}
An algorithm for generating a $u(t)$ from $H(j\omega)$, which is obtained from the proposed method,  is discussed in this section. For brevity, the case for generating a $2\times 1$ input is illustrated. The extension for generating a $r\times 1$ input is presented in Figure \ref{fig:input_design}. Conventionally, the input is generated as a sample path of a stochastic process with the desired power spectral density. The same principle is adopted in this paper as well. To elaborate further, let
\begin{equation}\small 
H(j\omega) = \left[
\begin{array}{cc}
H_{11}(j\omega) & H_{12}(j\omega) \\
H_{21}(j\omega) & H_{22}(j\omega)  \end{array}
\right] \mbox{~~and~~}
\phi(j\omega) = \frac{1}{2}H(j\omega)H^*(j\omega) =  
\left[
\begin{array}{cc}
\phi_{11}(j\omega)& \phi_{12}(j\omega) \\
 \phi_{21}(j\omega) & \phi_{22}(j\omega)
 \end{array}
\right].
\label{eq:H_example}
\end{equation}
 Naturally, 
 \begin{equation}
\begin{array}{l}
\phi_{11}(j\omega)  =  \frac{1}{2}\left(\abs{H_{11}(j\omega)}^2+\abs{H_{12}(j\omega)}^2\right), 
\phi_{22}(j\omega)  =  \frac{1}{2}\left(\abs{H_{21}(j\omega)}^2+\abs{H_{22}(j\omega)}^2\right)\\
\hspace*{1cm} \phi_{12}(j\omega)  = \frac{1}{2}\left(H_{11}(j\omega)H_{21}^*(j\omega)+H_{12}(j\omega)H_{22}^*(j\omega)\right)= \phi_{21}(j\omega)^*.
\end{array}
\end{equation}
The goal is then to determine a real-valued random vector $u(t) = \left[u_1(t), u_2(t)\right]^\top$ such that 
\begin{equation}
\mathcal{R}_u(\tau) = \mathbb{E}\left\{u(t)u^{*}(t-\tau)\right\} = \frac{1}{2\pi}\intop^{\omega_c}_{-\omega_c} \phi_u(j\omega) e^{j\omega\tau} d\omega,
\label{eq:need1}
\end{equation}
where $ \mathbb{E}\{.\}$ denotes the expected value. To that end, consider the one sided frequency functions $i.e.$ defined for $\omega \geq 0$,
\begin{equation}\small 
\begin{array}{l} 
U_1(j\omega) = \sqrt{\pi}(H_{11}(j\omega)e^{j\Phi_1^{\omega}} + H_{12}(j\omega)e^{j\Phi_2^{\omega}}) \mbox{~and} \;  
U_2(j\omega) = \sqrt{\pi}(H_{21}(j\omega)e^{j\Phi_1^{\omega}} + H_{22}(j\omega)e^{j\Phi_2^{\omega}}),  
\end{array}
\label{eq:rv_def}
\end{equation}
where the families of random variables $\left\{\Phi_1^{\omega}, \omega \geq 0\right\}$ and $\left\{\Phi_2^{\omega}, \omega \geq 0\right\}$ are such that
\begin{equation}
\begin{array}{l}
\mathbb{E}\left\{e^{j\Phi_i^{\omega}}\right\} = 0, ~\forall \omega \geq 0, ~i = 1,2 \mbox{~~and~~} \mathbb{E}\left\{e^{j\Phi_1^{\omega_1}}e^{-j\Phi_2^{\omega_2}}\right\} =  0,~ \forall   \omega_1 ,\omega_2  \ge 0\\ 
\hspace*{1.5cm}\mathbb{E}\left\{e^{j\Phi_i^{\omega_1}}e^{-j\Phi_i^{\omega_2}}\right\} =  \delta(\omega_1-\omega_2),~  \forall \omega_1,\omega_2  \ge 0,~ i = 1,2.
\label{eq:conditions_rv}
\end{array}
\end{equation}
For the input signals have to be real-valued, extend $U_1$ and $U_2$ to the negative imaginary axis as:
\begin{eqnarray}
 U_1(-j\omega) = U_1^*(j\omega) 
{\rm ~~and~~}  U_2(-j\omega) = U_2^*(j\omega).
\label{NegFreqa}
\end{eqnarray}
Define the inputs as: 
\begin{eqnarray}
u_1(t) = \frac{1}{2\pi}\intop_{-\omega_c}^{\omega_c}U_1(j\omega)e^{j\omega t} d\omega \mbox{~~and~~}
u_2(t) = \frac{1}{2\pi}\intop_{-\omega_c}^{\omega_c}U_2(j\omega)e^{j\omega t} d\omega
\label{eq:inputs_from_fft}.
\end{eqnarray}
The claim is that the power spectral density of $\left[u_1(t), u_2(t) \right]$ is equal to $\phi_{u}(j\omega)$. For proving the claim, note that by using Fubini's Theorem \cite{nair2001functional}, 
\begin{eqnarray}
\mathcal{R}_{11}(\tau) & = & \mathbb{E}\left\{ u_1(t)u_1^*(t-\tau)\right\}\nonumber \\ 
& = &\mathbb{E}\left\{\frac{1}{2\pi}\intop_{-\omega_c}^{\omega_c}\frac{1}{2\pi}\intop_{-\omega_c}^{\omega_c}U_1(j\omega_1)U_1^*(j\omega_2)e^{j\omega_1t}e^{-j\omega_2(t-\tau)}d\omega_1d\omega_2\right\} \nonumber \\ 
& = &\frac{1}{2\pi}\intop_{-\omega_c}^{\omega_c}\frac{1}{2\pi}\intop_{-\omega_c}^{\omega_c}\mathbb{E}\left\{U_1(j\omega_1)U_1^*(j\omega_2)\right\}e^{j\omega_1t}e^{-j\omega_2(t-\tau)}d\omega_1d\omega_2
\label{eq:proof_input_design}
\end{eqnarray}
From \eqref{eq:rv_def} and \eqref{eq:conditions_rv}, it can be deduced that 
\begin{equation}
\mathbb{E}\left\{U_1(j\omega_1)U_1^*(j\omega_2)\right\} = \pi \left(\abs{H_{11}(j\omega_1)}^2+\abs{H_{12}(j\omega_2)}^2\right)\delta(\omega_1-\omega_2).
\label{eq:exp_U1_U1star}
\end{equation}
Therefore,
\begin{eqnarray}
\mathcal{R}_{11}(\tau) = \frac{1}{2\pi}\intop_{-\omega_c}^{\omega_c}\frac{1}{2}\left(\abs{H_{11}(j\omega)}^2+\abs{H_{12}(j\omega)}^2\right)e^{j\omega \tau}d\omega.
\end{eqnarray}
On similar lines, it follows that
\begin{equation}
\begin{array}{l}
\hspace*{0.8cm} \displaystyle \mathcal{R}_{22}(\tau) = \frac{1}{2\pi}\intop_{-\omega_c}^{\omega_c}\frac{1}{2}\left(\abs{H_{21}(j\omega)}^2+\abs{H_{22}(j\omega)}^2\right)e^{j\omega \tau}d\omega \mbox{~~and}\\
\displaystyle\mathcal{R}_{12}(\tau) = \frac{1}{2\pi}\intop_{-\omega_c}^{\omega_c}\frac{1}{2}\left(H_{11}(j\omega)H_{21}^*(j\omega)+H_{12}(j\omega)H_{22}^*(j\omega)\right)e^{j\omega \tau}d\omega.
\end{array}
\label{eq:final1}
\end{equation}
It is now easy to note from the relation \eqref{eq:need1} applied to \eqref{eq:final1} that the power spectral density of $[u_1(t) u_2(t)]^{\top}$ is equal to $\phi(j\omega) = H(j\omega)H^*(j\omega)$. Now, a closed form expression for the integrals in  \eqref{eq:inputs_from_fft} is difficult to obtain and therefore it needs to be evaluated numerically by discretization of the frequency band. To this end, choose $\Phi_1^{\omega}$ and $\Phi_2^{\omega}$ to be independent Gaussian white noise vectors. Since $U_1(j\omega)$ is zero outside the interval $\left[-\omega_c, \omega_c\right]$,
\begin{eqnarray}
u_1(t) &=& \frac{1}{2\pi}\int^{\omega_c}_{-\omega_c} U_1(j\omega) e^{j\omega t} d\omega.
\label{Disc1}
\end{eqnarray}
With a positive integer $N$ such that $\frac{1}{N} \ll 1$, a Riemann sum  approximation of \eqref{Disc1} would be
\begin{eqnarray}
\tilde{u_1}(t)  =  \frac{1}{2\pi N}\sum^{N}_{k=-N} U_1\left(j\frac{k}{N}\right) e^{j\frac{k}{N} t}. 
\label{Disc4}
\end{eqnarray}
In a similar way, $\tilde{u_2}(t)$ can also be defined. The signal $\tilde{u}(t) = [\tilde{u_1}(t) ~\tilde{u_2}(t)]$ is then used for simulations and computations. For a general $r$-input system, the procedure has been shown in Figure \ref{fig:input_design}.

\begin{figure}[t]
\centering
\fbox{\begin{minipage}{34em}
\begin{equation}
\begin{array}{l}
\mbox{Generate independent white random vectors~} \Psi_1, \Psi_2, \dots,  \mbox{~and~} \Psi_r \\ \\
\displaystyle \mbox{Define~} U_i\left(j\frac{k}{N}\right)  =  \sqrt{\pi} \sum_{m=1}^{r}H_{m}\left(j\frac{k}{N}\right)e^{j\Psi_m(k)},~\forall k=1,\dots,N, ~\forall i=1,\dots,r. \\ \\
\displaystyle \mbox{Compute~}\tilde{u_i}(t) = \frac{1}{2\pi N}\sum^{N}_{k=-N} U_i\left(j\frac{k}{N}\right) e^{j\frac{k}{N} t}, \,\, \forall k=1,\dots,r, \,\, \forall i=1,\dots,r.
\label{eq:General_r_input}
\end{array}
\end{equation}
\end{minipage}}
\caption{Input Design Methodology.}
\label{fig:input_design}
\end{figure}

\end{document}